\documentclass{jpsj-suppl}
\usepackage{txfonts} %Please comment out this line unless the txfonts package is available in your LaTeX system.
\usepackage{epstopdf}

\title{Generators for the SIS/DIS region}

\author{Christophe \textsc{Bronner}$^{1}$}

\inst{$^{1}$Kavli Institute for the Physics and Mathematics of the Universe (WPI), The University
of Tokyo Institutes for Advanced Study, University of Tokyo, Kashiwa, Chiba, Japan}

\email{christophe.bronner@ipmu.jp}

\recdate{January 28, 2016}

\abst{We describe how the main neutrino interaction generators (GENIE, NEUT and NuWro) used by current neutrino oscillation experiments treat the shallow and deep inelastic region. We then compare their predictions for charged current events in this region, in terms of transferred momentum as well as multiplicities for different types of hadrons. We present additional comparisons in the low hadronic invariant mass region, where the generators use different custom models.}

\kword{Neutrino interaction generator, deep-inelastic, shallow-inelastic, generator comparisons}

\begin{document}
\maketitle

\section{Introduction}
Neutrino oscillation experiments rely on predictions obtained by Monte Carlo simulations to analyze their data. A crucial requirement for those simulations is the ability to properly simulate the interactions of neutrinos and anti-neutrinos with the nuclei of the elements constituting the detectors, to generate the particles which will then be propagated through a simulation of the detector to produce the final observables. This step is handled by neutrino interaction generators, and in this article we look at how the generators most commonly used by the current experiments handle the shallow and deep inelastic (SIS-DIS) region. We will focus on the case of the charged current (CC) interactions, and define this region as comprising all the events for which the invariant mass of the hadronic system W is larger than 1.7 GeV/c$^{2}$, for reasons which will be detailed in the next section.\\

\noindent We considered the following three neutrino interaction generators:
\begin{itemize}
\item NEUT\cite{NEUT} version 5.3.4
\item GENIE\cite{GENIE} version 2.10, comparing the predictions of the two models for final state interactions (FSI) hA and hN when relevant
\item NuWro\cite{NuWro} version 11q
\end{itemize}
Additionally, predictions of the GiBUU generator taken from \cite{GiBUU} will be shown on comparisons when possible.
When comparing the output of different generators, all the simulated events will be interactions of muon neutrinos and anti-neutrinos.

\section{Modelling of the shadow and deep inelastic region in the generators}\label{Models}
Above the pion production threshold, the three generators use different models depending on the value of the invariant mass of the hadronic system for the event to generate. The general picture is that at low W a combination of exclusive resonance channels and of a continuous DIS background is used, while at high W the predictions of the PYTHIA\cite{PYTHIA, PYTHIA6} generator are used. The number of resonances considered, as well as how the transition between those two regimes is done differ from generator to generator.\\

\paragraph{NEUT:} For W$<$2 GeV/c$^{2}$, resonance channels producing one pion, one kaon and one eta are considered, while the DIS background is handled by the `multi-pi' mode. This mode contains all the events in which 3 or more hadrons are produced. One of those hadrons is a nucleon, while all the remaining ones are assumed to be pions. The multiplicities are generated using a model based on Koba-Nielsen-Olesen (KNO) scaling\cite{KNO}. For W$>$2 GeV/c$^{2}$, no resonance channels are considered, and the predictions of PYTHIA 5.72 are used.\\

\paragraph{GENIE:} In the default version of GENIE, a larger number of resonances plus a continuous DIS background (the `low W AGKY model'\cite{AGKY}, also based on KNO scaling for the multiplicities of hadrons) are used for W$<$1.7 GeV/c$^{2}$. Between 1.7 and 2.3  GeV/c$^{2}$, no resonances are considered and only the low W AGKY model is used. A linear transition as a function of W is then done between this model and the predictions of PYTHIA6, which handles all the events for W$>$3 GeV/c$^{2}$.\\

\paragraph{NuWro:} In the case of NuWro, only resonances are considered for W$<$1.3 GeV/c$^{2}$. A linear transition as a function of W is done between those resonances and a DIS model which is the only model used above 1.6 GeV/c$^{2}$. This DIS mode uses fragmentation routines from PYTHIA6, with modifications so that the model can be used for lower values of W than PYTHIA can.\\

\begin{figure}[tbh]
\begin{center}

 \includegraphics[width=0.6\textwidth,clip]{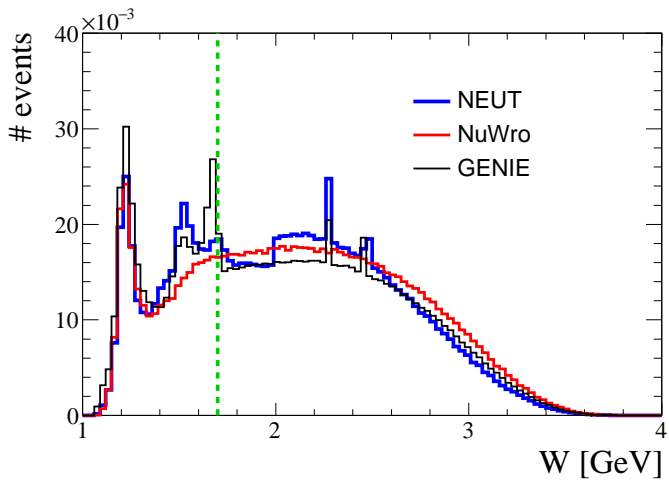} 

\caption{Distributions of the hadronic invariant mass of events generated by NEUT, NuWro and GENIE, corresponding to interactions of 6 GeV neutrinos on iron. Only the charged current deep inelastic and resonant modes of the generators are used. The dashed green line correspond to the boundary W=1.7 GeV/c$^{2}$ used to define the SIS-DIS region in this article.}
\label{Fe_W}
\end{center}
\end{figure}

All three generators use to compute the cross sections for their DIS modes the GRV98 parton distribution functions (PDF) \cite{GRV98} with corrections by Bodek and Yang\cite{BodekYang}. However the differences in the treatment of this region are very apparent on the W distributions of events simulated for 6 GeV neutrinos interacting with iron (figure \ref{Fe_W}).

 We can notice in particular differences in the number of peaks corresponding to the differences in the number of exclusive resonant channels taken into account by the different generators. As we aim at comparing the predictions of the generators for the DIS component and not the resonant modes, the threshold for events to be used in comparisons was set at W=1.7 GeV/c$^{2}$. This selects a region in which the events are mainly produced by the DIS modes of the generators, while also including the multi-pi mode from NEUT. We can also notice that the linear transitions allow for a smooth transition between the models, whereas simple change of model at a given W can lead to discontinuities in the W distributions. 

\section{Charged hadron multiplicities for interactions on free nucleons}\label{MultD2}
The generators turn the available W into a set of particles, mainly hadrons. We will first look at how many particles get created in this hadronization process. In a detector using massive elements as target material, the primary hadrons produced by the interaction of a neutrino and a nucleon could re-interact before they exit the nucleus, and in this case the multiplicity observed in the detector will be different from the multiplicity at the interaction level. To disentangle real multiplicity differences from differences in the treatment of final state interactions (FSI), we will compare in this section the multiplicities obtained in the case of interactions on free nucleons. Deuterium bubble chamber experiments measured the multiplicities of charged hadrons in neutrino and anti-neutrino interactions\cite{FNAL,BEBC}. They observed that the average charged hadron multiplicity was a linear function of the logarithm of W$^{2}$, and that those multiplicities were different for interactions of neutrinos and anti-neutrinos, as well as for interactions with a neutron and with a proton. In this section we will consider that deuterium is approximatively equivalent to free neutrons and protons, and compare the predictions of the generators to the results of the charged hadron multiplicities measured by deuterium bubble chamber experiments.\\

\begin{figure}[tbh]
\begin{center}
\begin{minipage}{1\textwidth}
 \includegraphics[width=0.5\textwidth,clip]{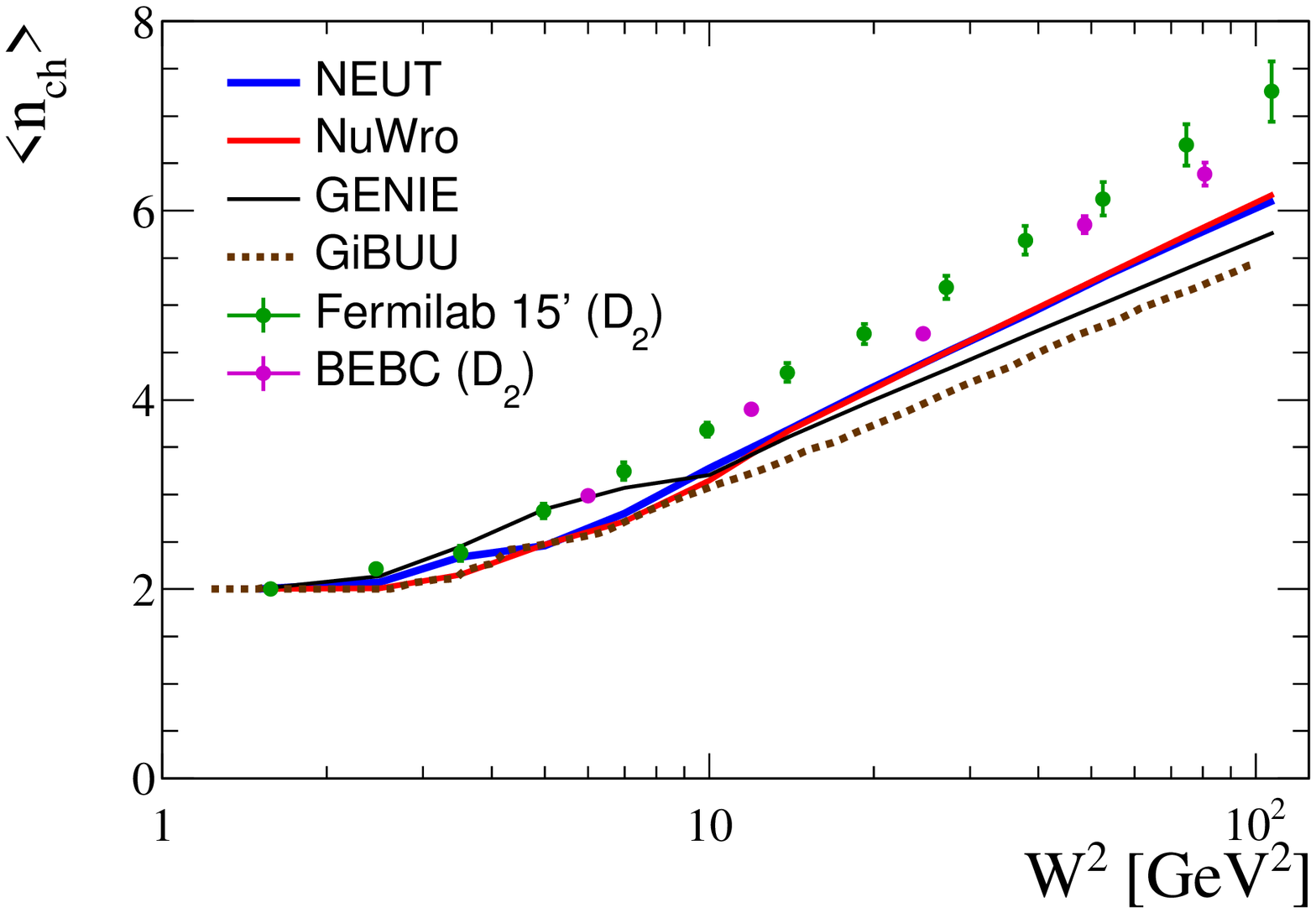} 
 \hfill
 \includegraphics[width=0.5\textwidth,clip]{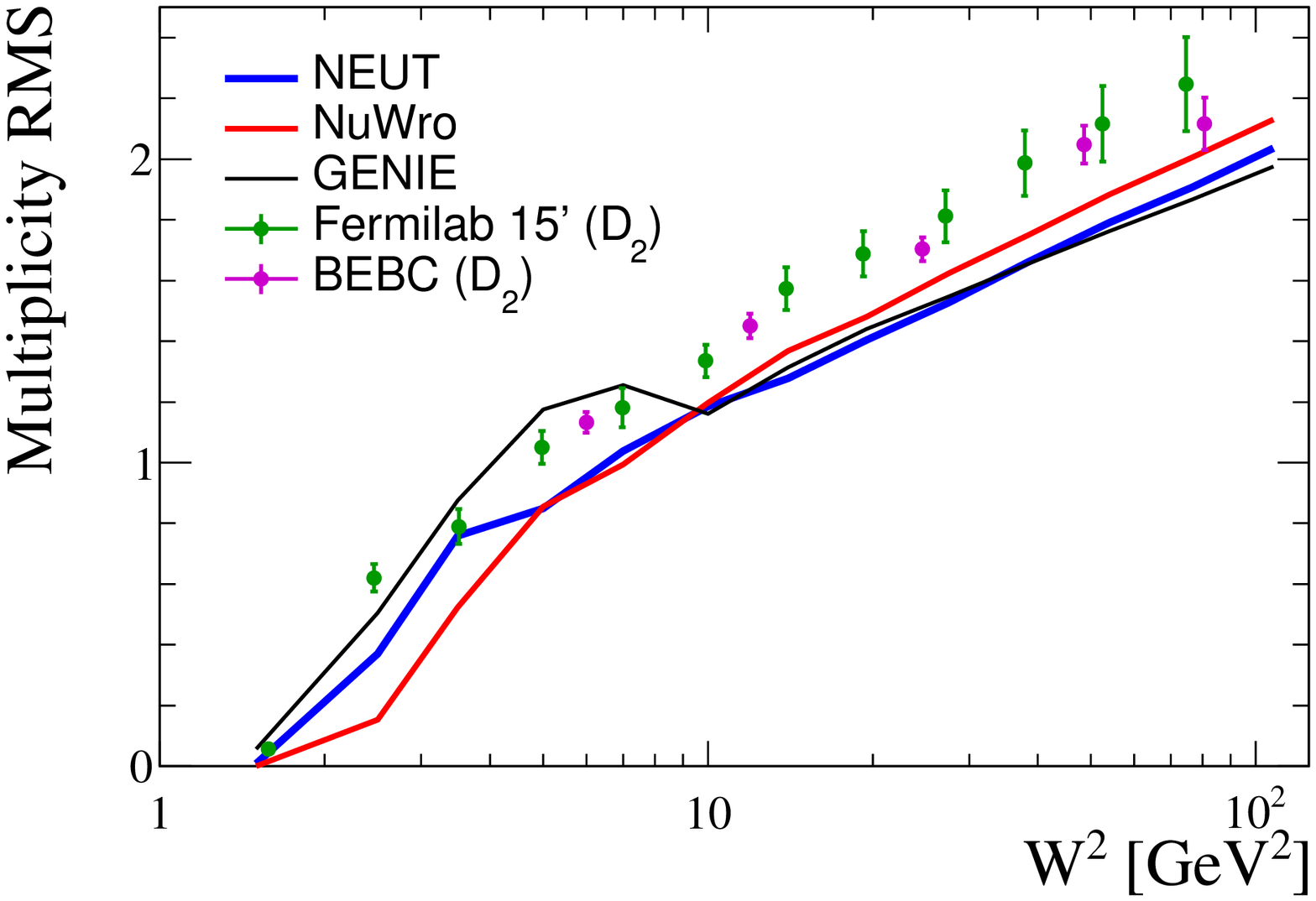}
\caption{Mean value (left plot) and dispersion (right plot) as a function of the square of the hadronic invariant mass W$^{2}$ of the number of charged hadrons produced in the interactions of neutrinos with protons.}
\end{minipage}
\label{MultNuP}
\end{center}
\end{figure}

To generate the events used in the comparisons, settings very similar to what is used in the validation tools for hadronization in GENIE were used:
\begin{itemize}
\item For GENIE and NuWro, free protons and neutrons are used as target
\item as NEUT does not allow directly to generate interactions on free nucleons, CH is used as a target, with FSI and nuclear effects turned off
\item the neutrino/anti-neutrino flux is inversely proportional to the neutrino energy on the range 0.5-80 GeV
\end{itemize}
Results obtained in the case of interactions of neutrinos on protons are shown on figure \ref{MultNuP}.

We can see that with the exception of GENIE in the region 5 GeV/c$^{2}<$ W$^{2}<$ 10 GeV/c$^{2}$, the different generators display similar behavior, and underestimate both the average and the dispersion of the number of charged hadrons produced in those interactions (although it has to be noted that the bubble chamber data used correspond to interactions on deuterium and not on free nucleons). Similar patterns were observed for the interactions of neutrinos on neutrons, and for the interactions of anti-neutrinos, although in the case of anti-neutrinos the difference are less significant as the data is scarce. 

There have been some attempts at tuning the hadron multiplicities in neutrino interaction generators, to obtain better agreement with the measurements from bubble chamber experiments. In \cite{Teppei}, the authors have tuned the parameters of the PYTHIA generator so that the average charged hadron multiplicity obtained with GENIE matches the average multiplicities measured by the bubble chamber experiments. They note however difficulties in reproducing the dispersion of the charged hadron multiplicity, as well as the multiplicities of neutral hadrons. The author of this paper presented a different method in a poster at the present conference, where a weight is assigned to the events generated by NEUT as a function of the hadron multiplicity and W of each event. This allows to reproduce the results of bubble chamber experiments for both the average value and the dispersion of the number of charged hadrons, but the effects of this tuning on the neutral hadrons multiplicities have not been checked.

\section{Comparisons for different fixed energies and targets}
In this section, we compare the predictions of the generators for interactions on massive targets. Unlike in the previous section, differences in the modelling of FSI between the generators will have an effect on the comparisons. Four couples of targets and fixed energies were considered, corresponding to cases which could be observed in currently running or planned experiments:
\begin{itemize}
\item 2 GeV $\nu$/$\overline{\nu}$ on CH (6 bound protons, 6 bound neutrons, 1 free proton)
\item 2.5 GeV $\nu$/$\overline{\nu}$ on argon (18 bound protons, 22 bound neutrons, 0 free protons)
\item 4 GeV $\nu$/$\overline{\nu}$ on water (8 bound protons, 8 bound neutrons, 2 free protons)
\item 6 GeV $\nu$/$\overline{\nu}$ on iron (26 bound protons, 30 bound neutrons, 0 free protons)
\end{itemize}
As before, only the resonant and DIS charged current modes of the generators are used, and a W$>$1.7 GeV/c$^{2}$ cut is applied. All plots are normalized by area.

\subsection{Transferred momentum Q$^{2}$}
\begin{figure}[tbh]
\begin{center}
\begin{minipage}{1\textwidth}
 \includegraphics[width=0.5\textwidth,clip]{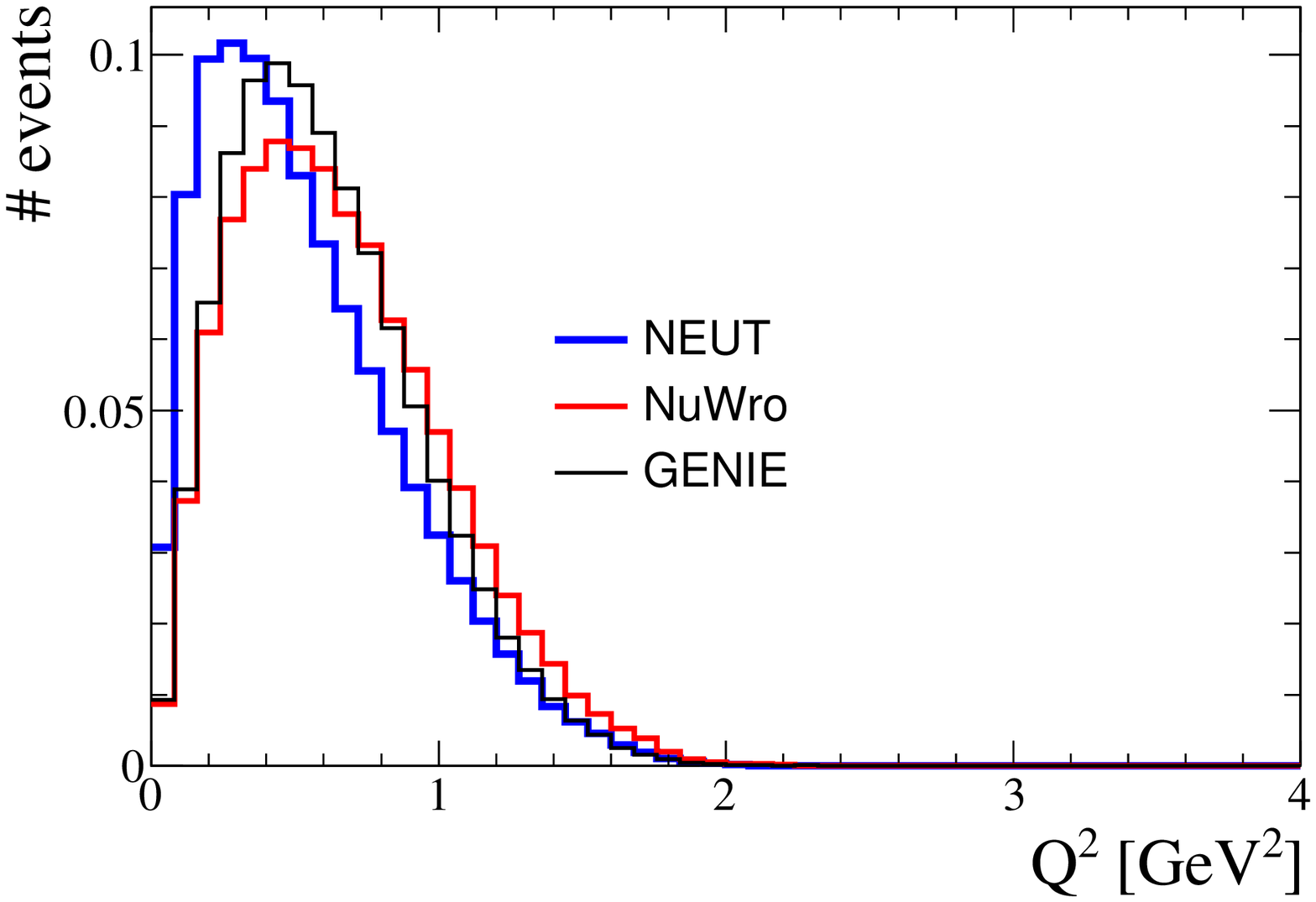} 
 \hfill
 \includegraphics[width=0.5\textwidth,clip]{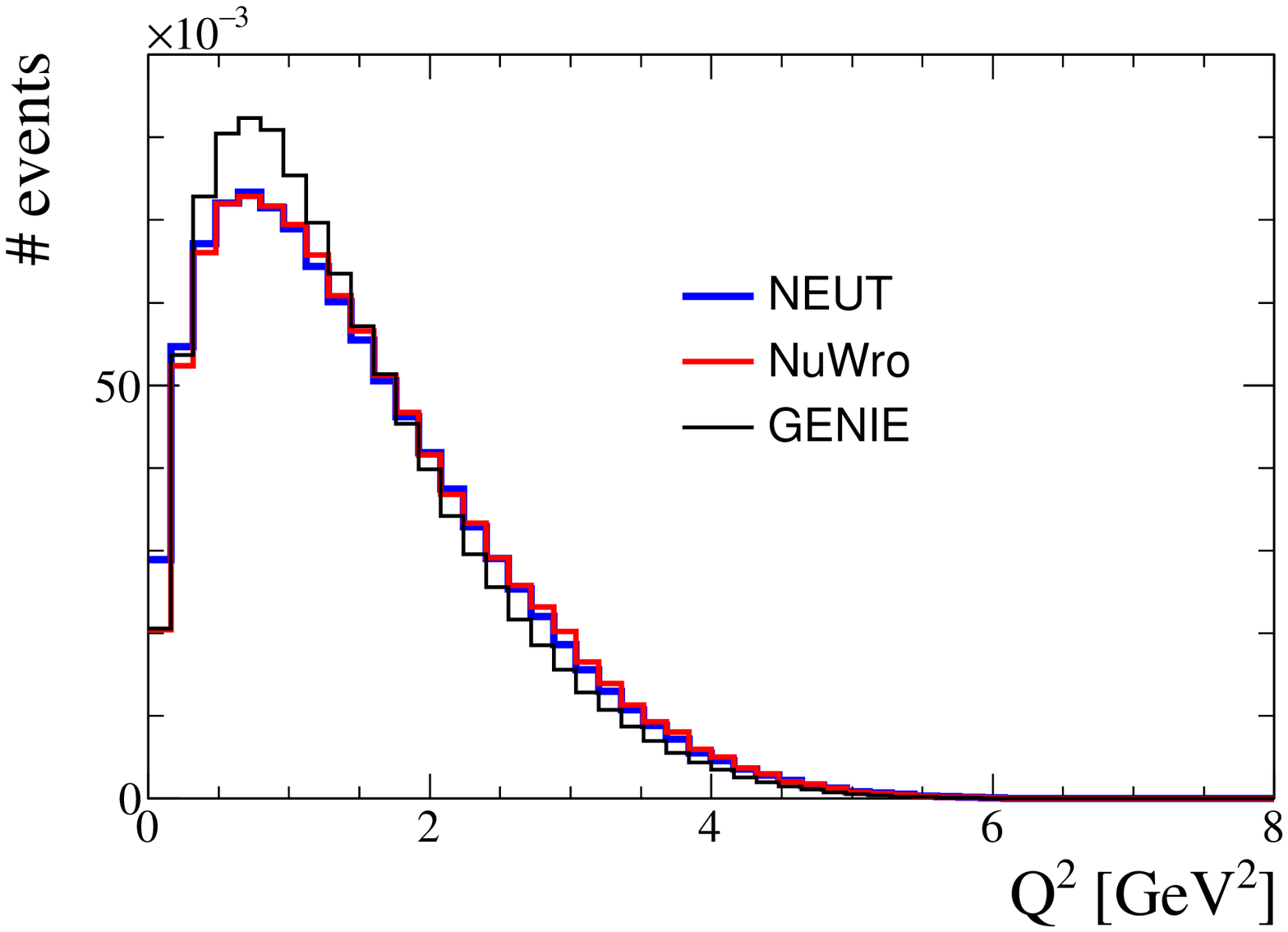}
\caption{Comparison of the transferred momentum distribution obtained with the three different generators for interactions of 2 GeV neutrinos on CH (left plot) and 4 GeV neutrinos on water (right plot).}
\end{minipage}
\label{NuQ2}
\end{center}
\end{figure}

\begin{figure}[tbh]
\begin{center}
\begin{minipage}{1\textwidth}
 \includegraphics[width=0.5\textwidth,clip]{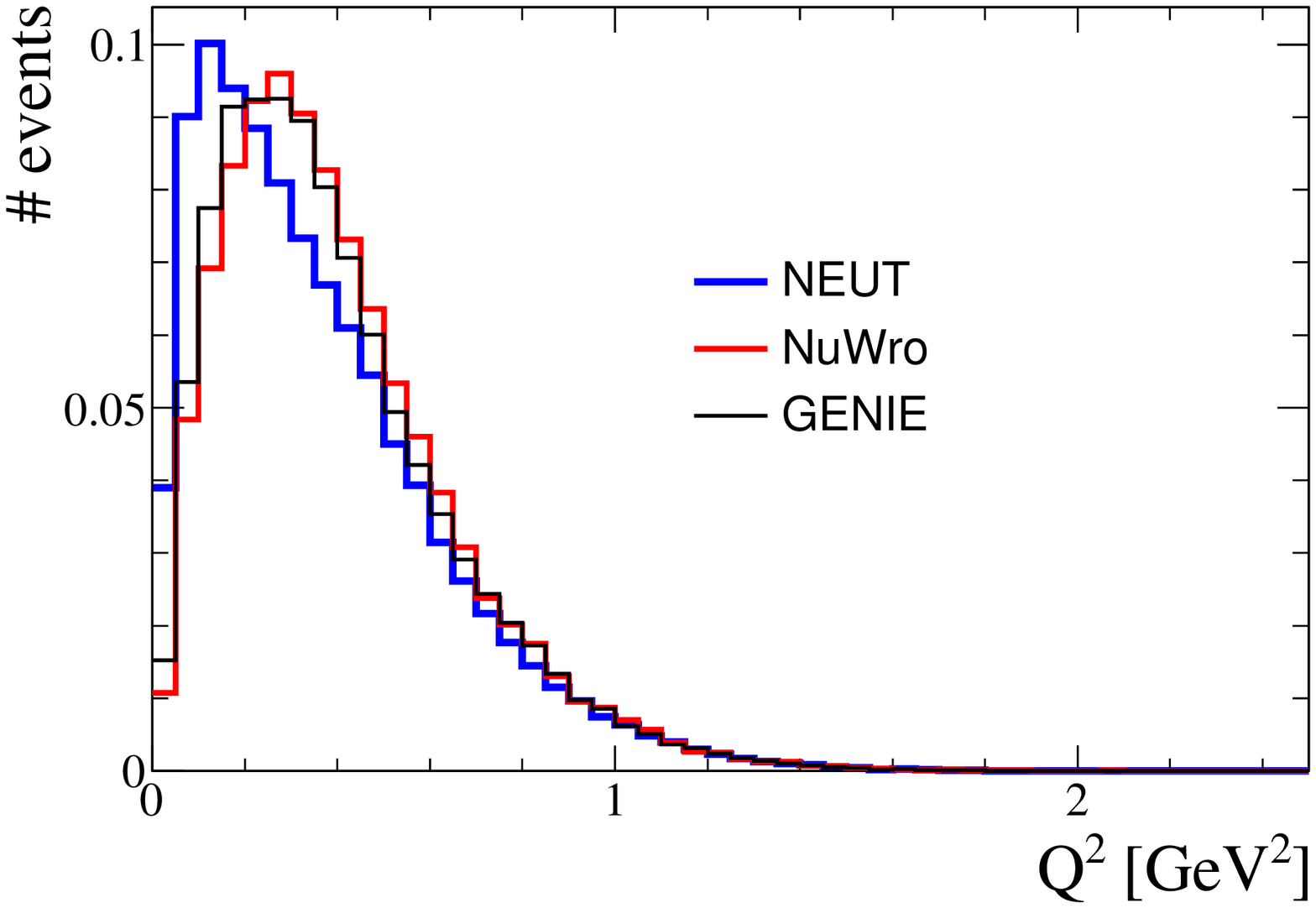} 
 \hfill
 \includegraphics[width=0.5\textwidth,clip]{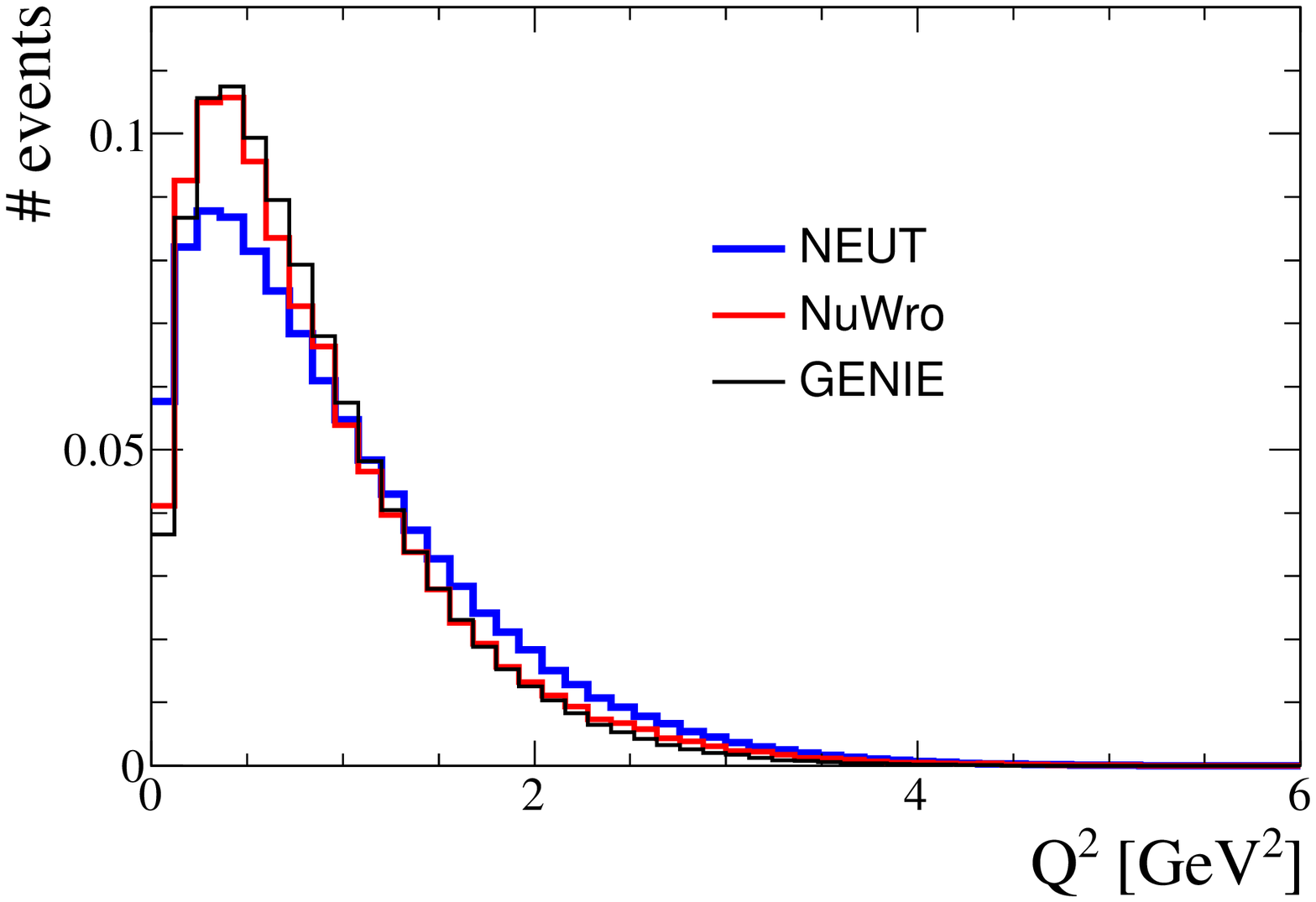}
\caption{Comparison of the transferred momentum distribution obtained with the three different generators for interactions of 2 GeV anti-neutrinos on CH (left plot) and 4 GeV anti-neutrinos on water (right plot).}
\end{minipage}
\label{NubarQ2}
\end{center}
\end{figure}

The first set of comparisons concerns the transferred momentum defined as $Q^{2}=(P_{\nu}-P_{lep})^{2}$ where $P_{\nu}$ is the four-momentum of the incoming neutrino or anti-neutrino, and $P_{lep}$ the four-momentum of the produced $\mu^{\pm}$. Although the three generators use similar PDFs to compute the cross sections for their DIS modes, differences in the Q$^{2}$ distributions of the events generated are seen for both neutrinos (figure \ref{NuQ2}) and anti-neutrinos (figure \ref{NubarQ2}). In the case of neutrino interactions, the distributions are peaked at similar values for GENIE and NuWro, but NuWro seems to generate more high Q$^{2}$ events. The behavior of NEUT is quite different from the other generators: its distribution is peaked at a lower transferred momentum for 2 GeV neutrinos, and is in very good agreement with NuWro at 4 GeV. For anti-neutrino interactions, there is a very good agreement between the Q$^{2}$ distributions of the events generated by GENIE and NuWro, while NEUT displays similar behavior as for neutrinos, with the peak of its Q$^{2}$ distribution being at a lower value for 2 GeV anti-neutrinos, and getting to values closer to the other generators for higher energies anti-neutrinos.

\subsection{Charged hadron multiplicities}
\begin{figure}[tbh]
\begin{center}
\begin{minipage}{1\textwidth}
 \includegraphics[width=0.5\textwidth,clip]{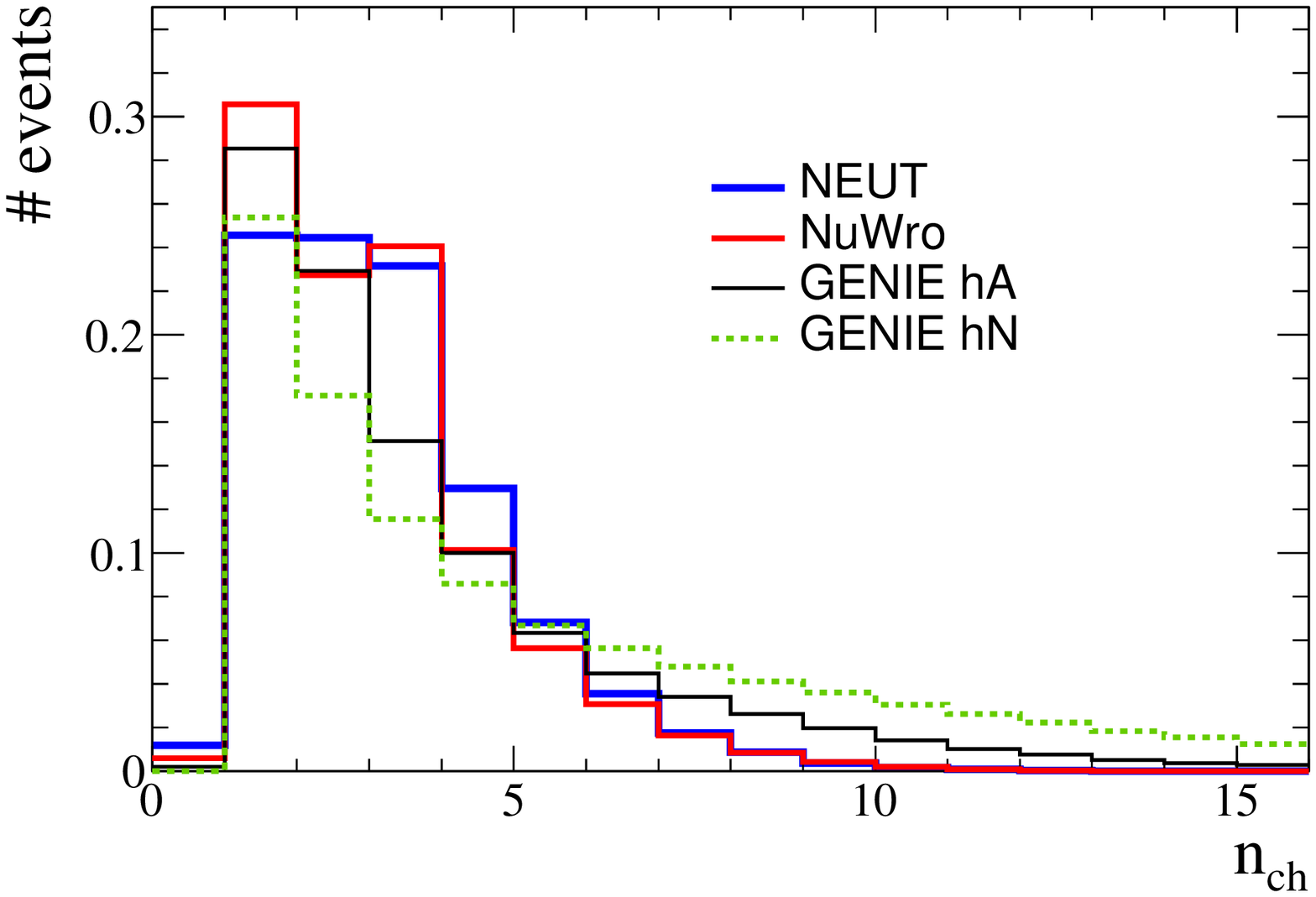} 
 \hfill
 \includegraphics[width=0.5\textwidth,clip]{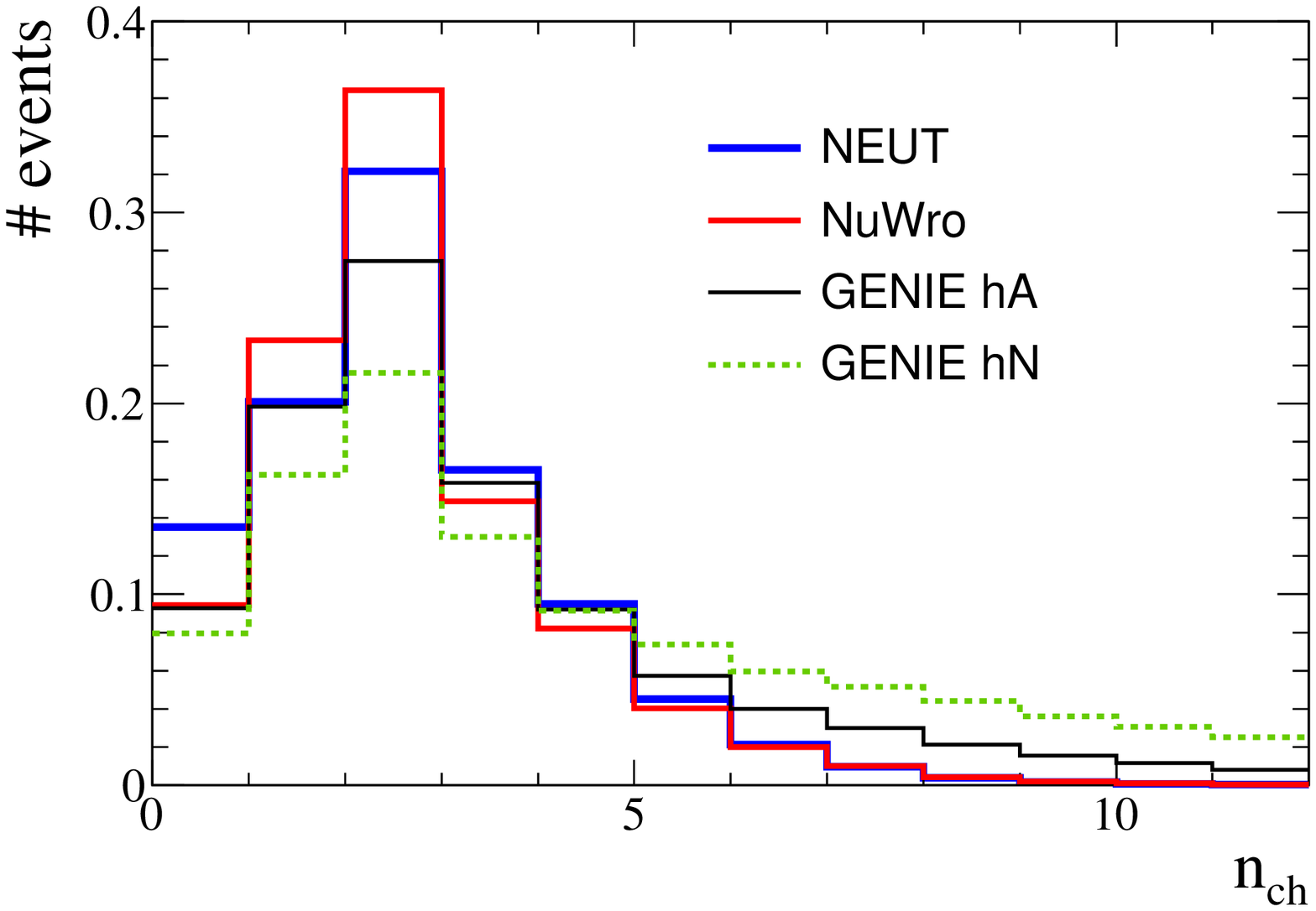}
\caption{Comparison of the number of charged hadrons in the events generated by the three generators for interactions of 2.5 GeV neutrinos (left plot) and anti-neutrinos (right plot) on argon.}
\end{minipage}
\label{ArNch}
\end{center}
\end{figure}

The number of charged hadrons produced in the interactions were compared for the four couples of targets and fixed energies. The case of 2.5 GeV neutrinos and anti-neutrinos on argon is shown on figure \ref{ArNch}. The main difference with section \ref{MultD2} is that we are no longer using free nucleons as target, and so the differences in the treatment of the re-interactions in the nucleus bring additional differences between the multiplicities predicted by the three generators. It was observed that GENIE predicted more charged hadrons than the other generators, and that the GENIE FSI model hN was producing events with higher multiplicities than the hA one. Although differences were seen between the predictions of the three generators for all couples of targets and energies, no other clear tendencies were observed.

\subsection{Pion multiplicities}
Similar comparisons were done for the multiplicities of pions, first looking at the total pion multiplicities, and then at the neutral pion multiplicities. The first observation was that the two GENIE FSI models were producing events with quite similar pion multiplicity distributions. In the case of neutrinos, NuWro was found to predict more pions at low energy, while the predictions of the different generators were becoming more similar when the energy of the neutrinos was increasing. For anti-neutrinos, NuWro predicted slightly more pions than the two other generators at all energies, and the predictions from NEUT and GENIE were getting closer and closer as the energy of the anti-neutrinos was increasing from 2 GeV to 6 GeV. 

When the comparisons were done for neutral pions, similar patterns were observed for neutrinos and anti-neutrinos. At 2 GeV, the predictions of GENIE and NEUT were very close, while NuWro was generating events with slightly more neutral pions. As the energy increased, the distributions for the events generated by NEUT were observed to move progressively away from the GENIE distributions and closer to the NuWro ones.

\section{Low W models}
PYTHIA cannot be used at low W, and so the generators use their own custom models to generate events in this region, as detailed in section \ref{Models}. Since the types and kinematics of the particles produced are assigned differently by the models of the different generators, we have run additional comparisons for those modes. The events were generated using the same settings as used to compare the multiplicities for interactions on free nucleons in section \ref{MultD2}, and a selection cut 1.7 GeV/c$^{2}<$ W$^{2}<$ 2 GeV/c$^{2}$ was applied to be in a domain where all three generators use their low W models.

\subsection{Leading pion momentum}
\begin{figure}[tbh]
\begin{center}
\begin{minipage}{1\textwidth}
 \includegraphics[width=0.5\textwidth,clip]{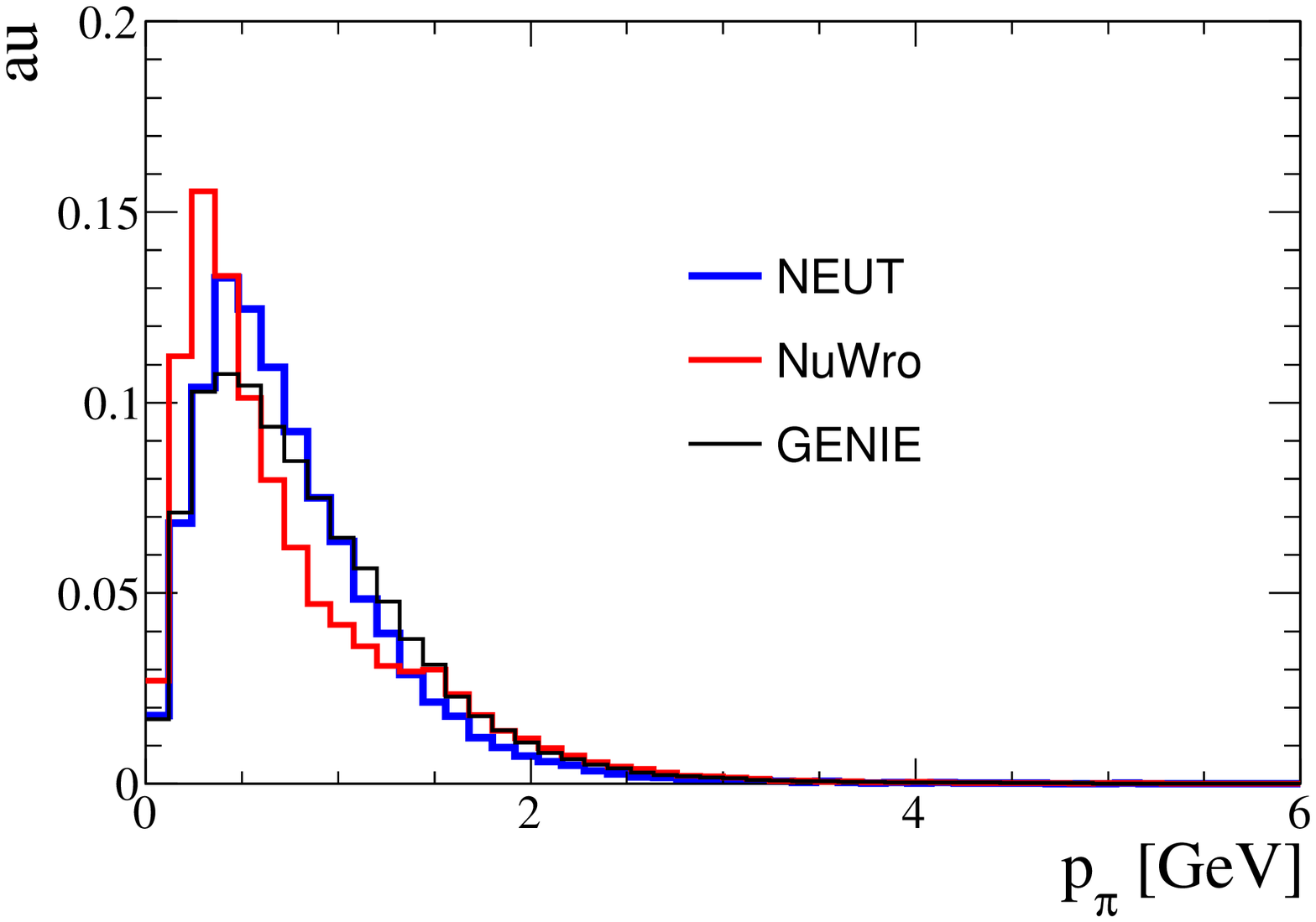} 
 \hfill
 \includegraphics[width=0.5\textwidth,clip]{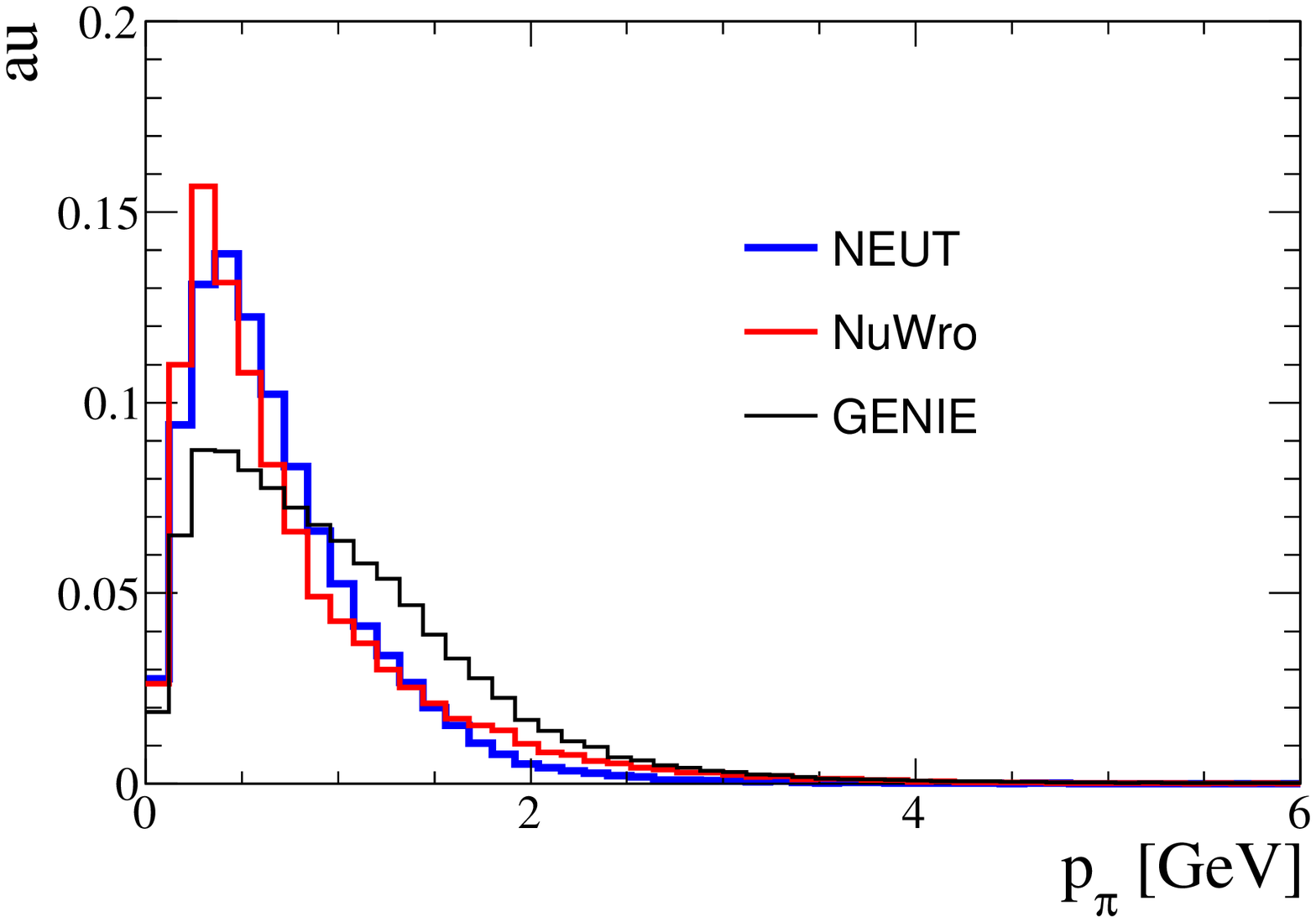}
\caption{Comparison of the leading pion momentum for neutrino events generated by the low W models of the three generators. The left plot correspond to interactions with protons, and the right plot to interactions with neutrons.}\label{fig:Pmom}
\end{minipage}

\end{center}
\end{figure}

The leading pion was defined as the $\pi^{+}$ with the largest momentum for neutrino interactions, and the $\pi^{-}$ with the largest momentum for anti-neutrino ones. The distributions were found to be different for interactions with protons and with neutrons, as a result the comparisons were done separately for each target nucleon. The results in the case of neutrino interactions are shown on figure \ref{fig:Pmom}. We can see that for interactions on protons, NuWro predicts lower momentum for the leading pion. The distributions from GENIE and NEUT are in relative agreement, although GENIE predicts more events with high pion momentum. In the case of interactions on neutrons, there is a good agreement between the predictions of NEUT and NuWro, while the spectrum is much broader for GENIE, which generates more events with large pion momentum. Similar patterns were observed for anti-neutrino interactions, when inverting the target nucleons compared to the neutrino cases. 

\subsection{Fraction of outgoing nucleons which are protons}
\begin{figure}[tbh]
\begin{center}
\begin{minipage}{1\textwidth}
 \includegraphics[width=0.5\textwidth,clip]{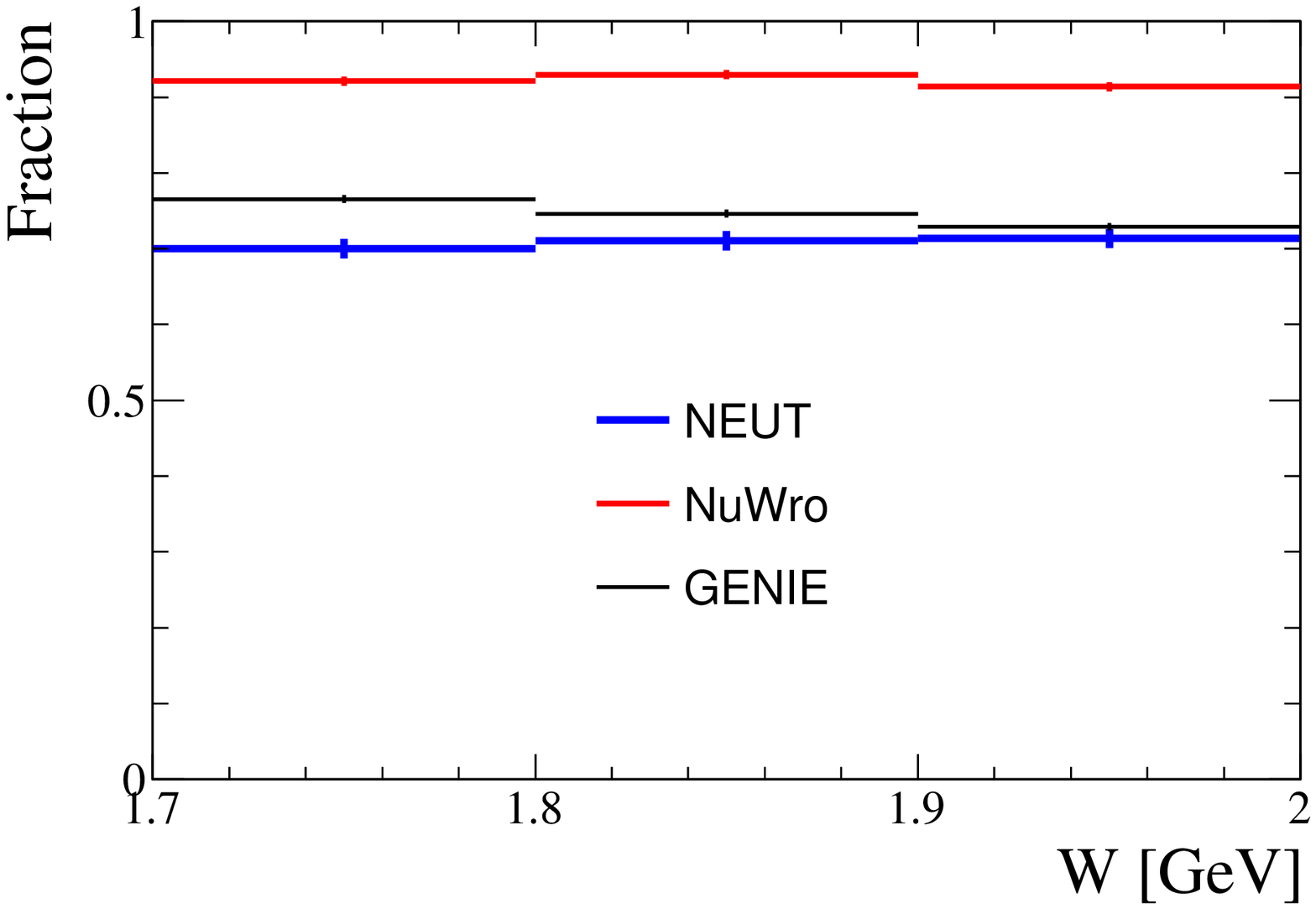} 
 \hfill
 \includegraphics[width=0.5\textwidth,clip]{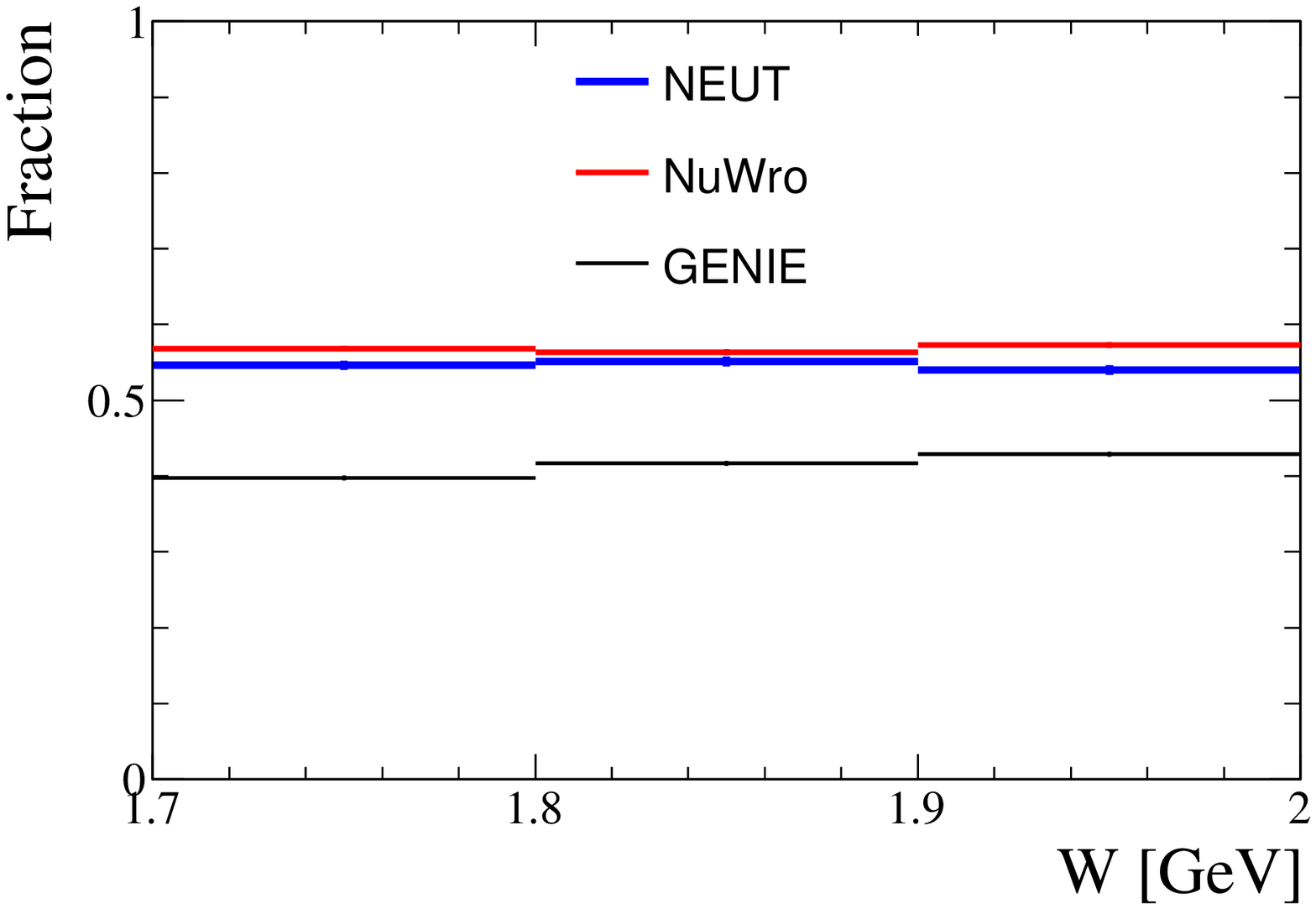}
\caption{Comparison of the fraction of outgoing nucleons which is a proton for neutrino events generated by the low W models of the three generators. The left plot correspond to interactions with protons, and the right plot to interactions with neutrons.}\label{fig:ProtonNu}
\end{minipage}

\end{center}
\end{figure}

\begin{figure}[tbh]
\begin{center}
\begin{minipage}{1\textwidth}
 \includegraphics[width=0.5\textwidth,clip]{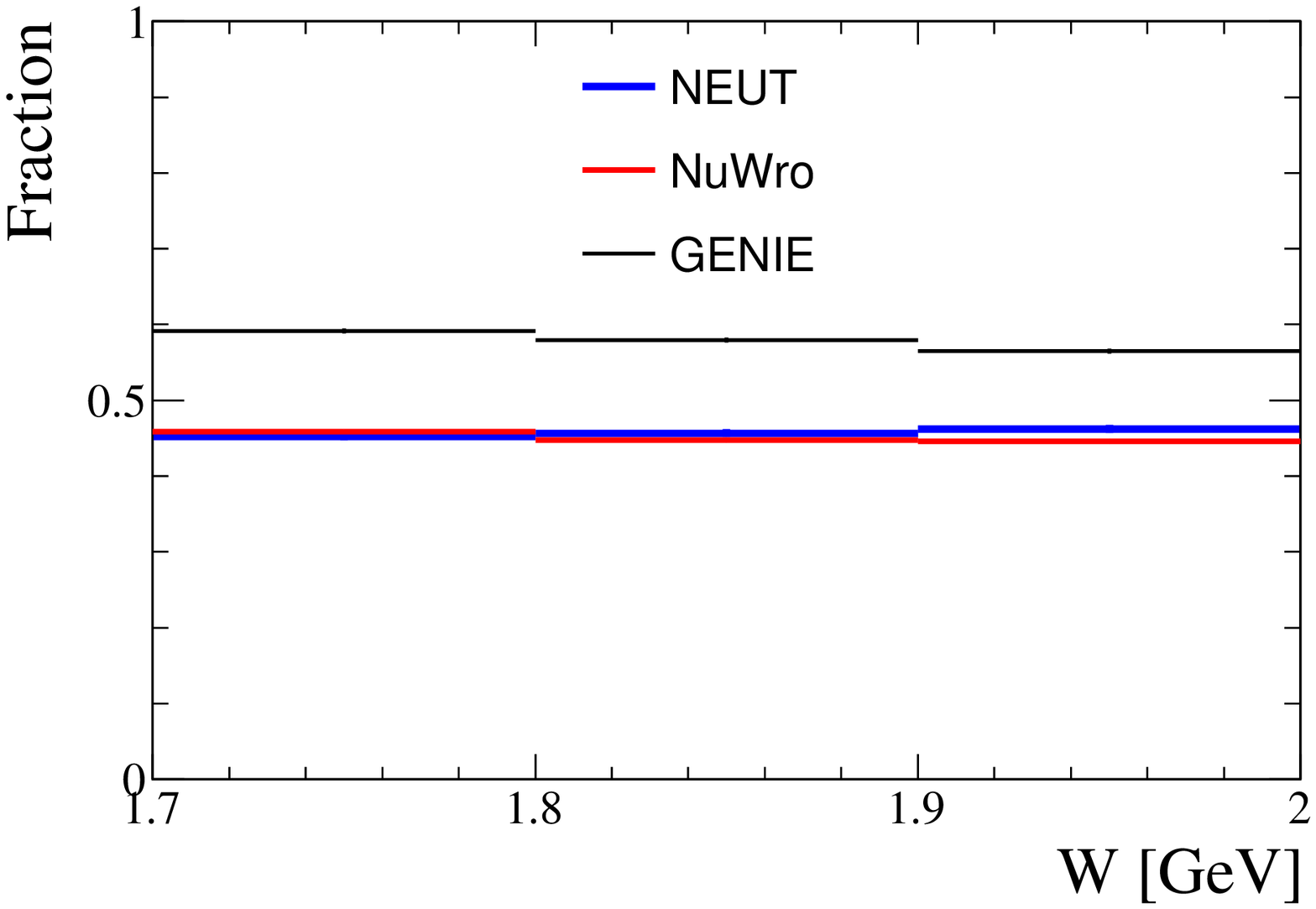} 
 \hfill
 \includegraphics[width=0.5\textwidth,clip]{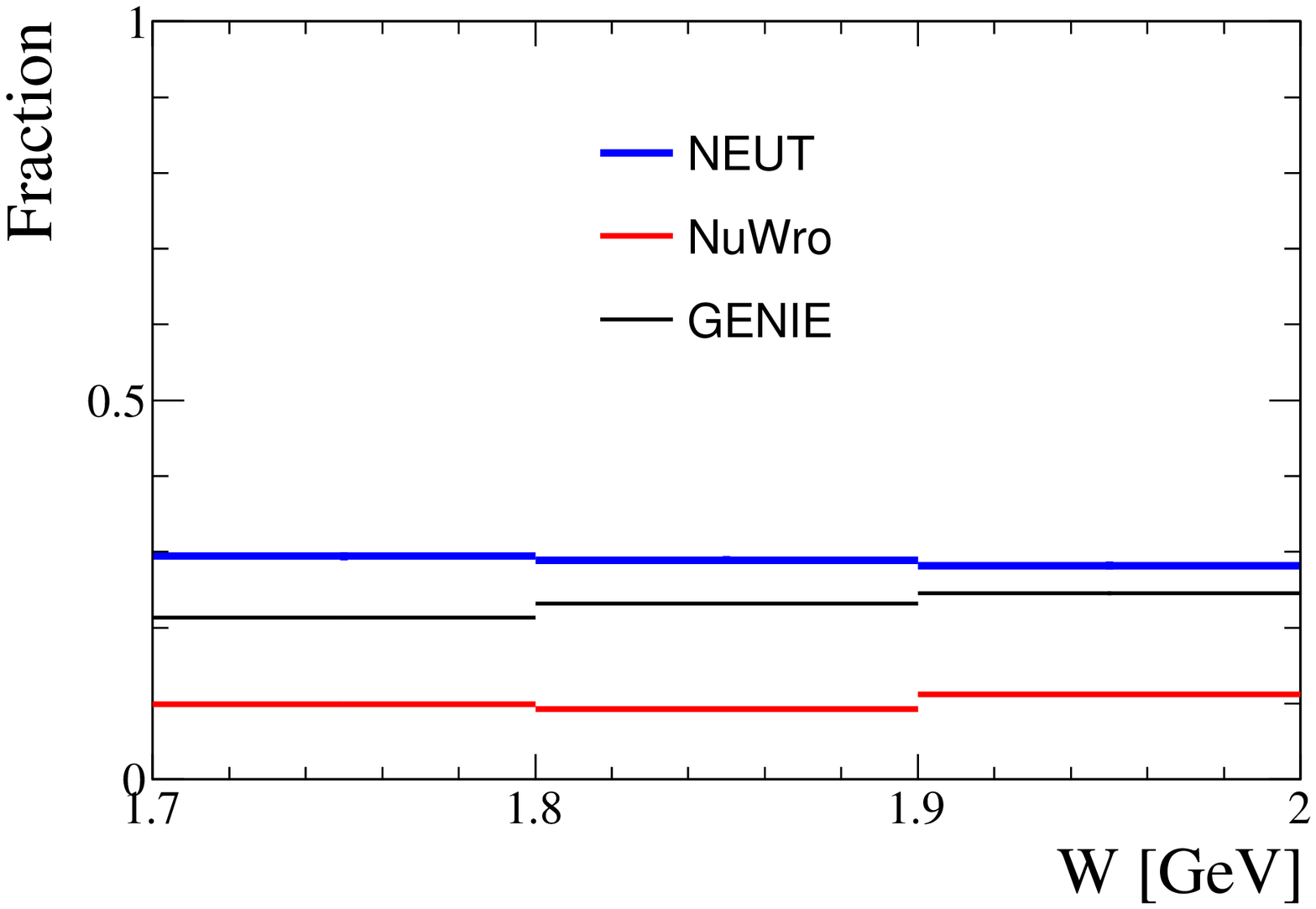}
\caption{Comparison of the fraction of outgoing nucleons which is a proton for anti-neutrino events generated by the low W models of the three generators. The left plot correspond to interactions with protons, and the right plot to interactions with neutrons.}\label{fig:ProtonNuBar}
\end{minipage}

\end{center}
\end{figure}

To conserve the baryonic number, at least one of the hadrons produced has to be a baryon. In this low W region, there is generally only one baryon produced, and it is most of the time a nucleon. We have looked at the fraction of those nucleons which are protons for events where only one baryon was produced, and this baryon was a nucleon. Those comparisons were done using the same events as the comparisons of the leading pion momentum. In this case as well the comparisons were done separately for the two possible target nucleons. The results for interactions of neutrinos are shown on figure \ref{fig:ProtonNu} and on figure \ref{fig:ProtonNuBar} for the interactions of anti-neutrinos. In all cases, the proton fractions appear to be quite independent of the hadronic invariant mass W. For $\nu$-p, $\nu$-n, and $\overline{\nu}$-p, two of the generators are in agreement while the third one has a different fraction of produced nucleons that are protons. In the case of $\overline{\nu}$-n, the three generators have different proton fractions.

\subsection{Pion fractions}
We have also looked at the fractions of the pions produced that were  $\pi^{+}$,  $\pi^{-}$ and  $\pi^{0}$ in those same events. The distributions were found to be different for interactions of neutrinos and anti-neutrinos, and for different target nucleons, and were very similar for GENIE and NEUT. The fractions for the events generated by NuWro were not too different from what was observed for the two other generators, but NuWro was producing more neutral pions than NEUT and GENIE.

\section{Summary}
We have described how the neutrino interaction generators NEUT, GENIE and NuWro were simulating events in the shallow and deep inelastic region, and compared properties of those events. It was found that all three generators were predicting a lower number of charged hadrons than what was measured by deuterium bubble chamber experiments. When comparing the predictions for different couples of fixed neutrino energies and targets, NEUT was found to have a different distribution of transferred momentum than the other generators. When looking at events generated with the models the generators use at low W, differences were seen in the leading pion momentum and in the fraction of the produced nucleons that were protons.

\end{document}